\documentclass[10.5pt,a4paper]{article}
\usepackage{amsmath}
\usepackage{ltablex}
\usepackage[subrefformat=parens]{subcaption}
\usepackage[margin=20pt]{subcaption}
\usepackage[labelformat=simple]{subcaption}
\usepackage{subcaption}
\usepackage[dvipdfmx]{graphicx}
\usepackage{booktabs}
\usepackage{here}
\usepackage[title]{appendix}

\usepackage{color}
\graphicspath{{Definitions/}} 
\usepackage{enumerate}
\usepackage{paralist}


\usepackage[top=30truemm,bottom=30truemm,left=20truemm,right=20truemm]{geometry}
\usepackage{fancyhdr}
    
\usepackage{amsmath}
\usepackage{amssymb}
\usepackage{mathcomp}
\usepackage{siunitx} 
\usepackage{cases}

\usepackage{authblk} 
\usepackage{color}

\usepackage{comment}

\title{Optimization of Design Parameters for Gravitational Wave Detector DECIGO Including Fundamental Noises}

\author[1]{Yuki Kawasaki}
\author[1]{Ryuma Shimizu}
\author[1]{Tomohiro Ishikawa}
\author[2]{Koji Nagano}
\author[1]{Shoki Iwaguchi}
\author[1]{Izyumi Watanabe}
\author[1]{Wu Bin}
\author[3,4]{Shuichiro Yokoyama}
\author[1,3]{Seiji Kawamura}

\affil[1]{\small{Department of Physics, Nagoya University, Nagoya, Aichi 464-8602}}
\affil[2]{\small{Institute of Space and Astronautical Science, Japan Aerospace Exploration Agency, \mbox{Sagamihara, Kanagawa 252-5210, Japan}}}
\affil[3]{\small{The Kobayashi-Masukawa Institute for the Origin of Particles and the Universe, Nagoya University, \mbox{Nagoya, Aichi 464-8602, Japan}}}
\affil[4]{\small{Kavli IPMU (WPI), UTIAS, The University of Tokyo, \mbox{Kashiwa, Chiba 277-8583, Japan}}}

\date{}

\begin{document}
\maketitle


\begin{abstract}
\noindent
 The DECi-hertz Interferometer Gravitational-Wave Observatory (DECIGO) is a space gravitational wave (GW) detector. DECIGO was originally designed to be sensitive enough to observe primordial GW background (PGW). However, due to the lowered upper limit of the PGW by the Planck observation, further improvement of the target sensitivity of DECIGO is required. In the previous studies, DECIGO’s parameters were optimized to maximize the signal-to-noise ratio (SNR) of the PGW to quantum noise including the effect of diffraction loss. To simulate the SNR more realistically, we optimize DECIGO's parameters considering the GWs from double white dwarfs (DWDs) and the thermal noise of test masses.  We consider two cases of the cutoff frequency of GWs from DWDs. In addition, we consider two kinds of thermal noise: thermal noise in a residual gas and internal thermal noise. To investigate how the mirror geometry affects the sensitivity, we calculate it by changing the mirror mass, keeping the mirror thickness, and vice versa. As a result, we obtained the optimums for the parameters {that} maximize the SNR that depends on the mirror radius. This result shows {that a thick mirror with a large radius gives a good SNR} and enables us to optimize the design of DECIGO based on the feasibility study of the mirror size in the future. 
\noindent

\end{abstract}

\newpage
\section{Introduction}
The DECi-hertz Interferometer Gravitational-Wave Observatory (DECIGO) is a space gravitational wave (GW) detector \cite{ref1,Kawamura}. One of the most important DECIGO goals is the observation of the primordial GW {background} (PGW) from the early Universe. It is ideal to observe the PGW in the lowest possible frequency band because the PGW has a larger strain in the lower frequency band. However, in the low-frequency band less than 0.1 Hz, GWs from {several astrophysical sources} impede the detection of the PGW  \cite{ref2}. Therefore, DECIGO, which has a frequency band between 0.1 Hz and 10 Hz, is optimized for the PGW~observation. 

 {Direct detection of PGW could} contribute to the determination of inflation models in the early Universe.
DECIGO was originally designed to be sensitive enough to observe the PGW, under the assumption that the normalized GW energy density $\Omega_{\mathrm{gw}}$ of the PGW is $\Omega_{\mathrm{gw}} \approx 2 \times 10^{-15} $ \cite{Kawamura2}. 
However, recent observations of CMB by Planck satellite and BICEP/Keck collaboration have lowered the upper limit for the PGW to $\Omega_{\mathrm{gw}} \approx 10^{-16} $, and this limit requires improvement of DECIGO’s sensitivity \cite{ref3, Planck2, ref4}.

In the previous studies, DECIGO’s parameters such as mirror reflectivity, arm length, and laser power were optimized for a given mirror radius to maximize signal-to-noise ratio (SNR) of the PGW to quantum noise, including the effect of diffraction loss \cite{Ishikawa, Iwaguchi}.
These parameters affect the magnitude of quantum noise in DECIGO. Specifically, the SNR increased from 6.6 to about 100 by the optimization.

The main noises {limiting the detection of PGW} are GWs from double white dwarfs (DWDs) and thermal noise.
DWDs are fast-rotating binary {stars} that emit GWs in the form of quadrupole radiation. 
(The frequency of the radiating GW is twice the angular frequency of the binary orbit.)
GWs from DWDs strongly affect the foreground GWs below around 0.1 Hz \cite{DWD}.
GWs from DWDs that cannot be resolved individually are regarded as noise.
{It is limiting} the detectable frequency band of the PGW.
In this paper, we estimate the limitation by GWs from DWDs.
In addition, limits to the detector thermal noise cannot be avoided because the mirror and its environment exhibit thermally-driven motion.
Thermal noise has larger effects on the SNR in the lower frequency band.

Due to the characteristics of GWs from DWDs and thermal noise, it is important to consider them to calculate the SNR to PGW.
Thus, we take GWs from DWDs and thermal noise into consideration.
Considering GWs from DWDs and thermal noise, we optimize DECIGO's parameters which are related to the magnitude of the noise and consider the parameters that give a larger SNR.

 In this paper, we show the noise by GWs from DWDs in Section \ref{GWs from DWDs}, the thermal noises in Section \ref{Thermal noises}, the method of optimization in Section \ref{Method of optimization}, the result of optimization in Section \ref{Results}, and the conclusion in Section \ref{Conclusion}.  
\section{GWs from DWDs \label{GWs from DWDs}}
GWs from various binaries are believed to affect the foreground GWs significantly in the frequency band between $10^{-5}$ and $10^{-1}$ Hz. 
The contribution of GWs from DWDs is especially relevant in the LISA band \cite{LISA1}. The contribution of galactic DWD is  {especially relevant} in that frequency band.
On the other hand, it is considered that the contribution of extragalactic DWD is important in the DECIGO band, which is discussed in e.g., Ref.~\cite{DWD}.
There are many DWDs in the Universe \cite{DWD}. When multiple GW signals from DWDs exist in one frequency bin, which is the frequency resolution of the detector, they cannot be resolved individually and are regarded as noise. 
A DWD binary system loses a part of energy by emitting GWs. The two stars of a DWD approach each other with increasing rotation frequency. 
Eventually, they collide at a certain frequency (called a cutoff frequency) which is determined by the mass and size of the binary star components. 
The DWD does not emit significant GWs after a collision. 
Therefore, the GWs from DWDs mainly exist in the frequency band below the cutoff frequency.

Due to their typical size and mass, DWDs emit GWs up to around 0.1 Hz. 
In the previous DECIGO design effort, the lower limit of the calculation range of the SNR was set to be 0.1 Hz to avoid the contamination from the foreground noise from DWDs below \mbox{0.1 Hz \cite{Ishikawa, Iwaguchi}}. 
On the other hand, most of the WDs observed near our galaxy have a mass of 0.8 $\mathrm{M_{\odot}}$ or less. 
If they form a binary star, they all coalesce at a frequency around \mbox{0.07 Hz \cite{ref2}}.
Thus, if the observed mass distribution of WDs is extended to the entire Universe, we can consider a model in which the noise from DWDs exists up to 0.07 Hz.
Therefore, in this paper, we consider two cases: for a lower limit of calculation of 0.1 Hz and 0.07 Hz. We decided to use these frequencies (0.1 Hz and 0.07 Hz) instead of providing two cases of the corresponding noise spectrums. 
We call these two cases the standard-DWD-model and the optimistic-DWD-model.
In Section \ref{Results}, we show the result of the calculation and optimized parameters for each model.

\section{Thermal Noise\label{Thermal noises}}
In this section, we estimate the noise caused by the thermal motion of DECIGO's mirrors.
We optimize the SNR of DECIGO for the PGW background, considering two sources of thermal noise: thermal noise in a residual gas and internal thermal noise.  
The former is caused by the collision of residual gas molecules, which thermally move in the satellite, with the mirror. 
We can calculate this from a simple model (see Section \ref{gasgas}). The latter comes from the internal dissipation of the mirror itself. 

Both power spectrums can be calculated using the Fluctuation-Dissipation Theorem. 
The theory states that the power spectrum of fluctuating displacement is given as \cite{fluctuation}
\begin{equation}
\label{FDT}
S_{x}(f)=\frac{k_{B}T}{\pi ^2 f^2}\mathrm{Re}\left[Y(f)\right].
\end{equation}

The function 
 $Y(f)$, called admittance, is 
\begin{equation}
v=Y(f)F_{ext},
\end{equation}
{where $v$ is the velocity of the mirror.}
In the following subsections, we use the Fluctuation-Dissipation Theorem to specifically calculate the thermal noise in a residual gas and the internal thermal noise.
Table \ref{tab1} shows the meaning of symbols used in this section.
\begin{table}[H] 
\centering
\caption{Meaning of each symbol.\label{tab1}}
\setlength{\tabcolsep}{11.2mm}\begin{tabular}{cc}
\toprule
\textbf{Symbol}	& \textbf{Meaning}	\\
\midrule
$k_{\mathrm{B}}$ &Boltzmann constant\\
$T$ (=300 {K})		&Mirror temperature\\
$m$		& Mirror mass	\\
$R$ &Mirror radius\\
$S$      &Mirror cross section\\
$h$ &Mirror thickness\\
$d$ &Coating thickness\\
$L$ &Cavity length\\
$E_0$ (=7.4 $\times$ 10$^{10}\ \mathrm{N/m^2})$ &Young's modulus\\
$\sigma$   (=0.17) & Poisson's modulus\\
$\alpha$ &Thermal expansion rate \\
$C$ &Specific heat per volume\\
$\kappa$ &Diffusivity of the mirror\\
$r_0$ &Beam radius\\
$P$ &Pressure in the satellite\\
$\mu$ &Mass of a gas molecule in the satellite\\

\bottomrule
\end{tabular}
\end{table}
\subsection{Thermal Noise in a Residual Gas\label{gasgas}}
The dissipation of the system is key in calculations using the Fluctuation Dissipation Theorem. 
First, we formulate the dissipation of the mirror due to the interaction between the mirror and its surroundings.
We model the mirror and its surrounding space \cite{Saulson}. 
The mirror with mass $m$ and cross-section $S$ floats in the satellite, and residual gas occupies its surroundings.

The pressure of the gas is $P = nk_{B}T$, where $n$ is the number density of gas molecules. 
We consider the case where $P$ is low enough so that the mean free path length of molecules is larger than the mirror size. 
In other words, we do not consider intermolecular collisions. 

Then, under this model, we obtain the power spectrum of the thermal noise in a residual gas $S_{x_{\mathrm{gas}}}(f)$,
\begin{equation}
\label{gas_power}
S_{x_{\mathrm{gas}}}(f)=\frac{k_{B}T}{\pi ^2 f^2}\frac{b}{(2\pi fm)^2 + b^2},
\end{equation}
where the factor $b$ is
\begin{equation}
\label{gas_factor}
b=2\pi \left[1.064\times 10^{13} \left(\frac{1\times 10^{-8}\ \mathrm{Pa}}{P}\right) \left(\frac{400\ \mathrm{cm^2}}{S}\right)\right]^{-1} \left[ \mathrm{kg/s}\right].
\end{equation}

In Equation (\ref{gas_power}), $b$ is an important factor that determines the scale of the power spectrum.
See Appendix \ref{appendixA} for how to derive $S_{x_{\mathrm{gas}}}(f)$ and $b$. 
In the calculation of $b$, it is assumed that the residual gas in the satellite is nitrogen.
If the main gas component is water instead, the coefficient of Equation (\ref{gas_factor}) should be $1.33 \times 10^{13}$ instead of $1.064\times 10^{13}$. 
Also, if the main gas component is hydrogen, the coefficient should be $3.98  \times 10^{13}$.

\begin{figure}[H]
\centering
\includegraphics[width=10cm]{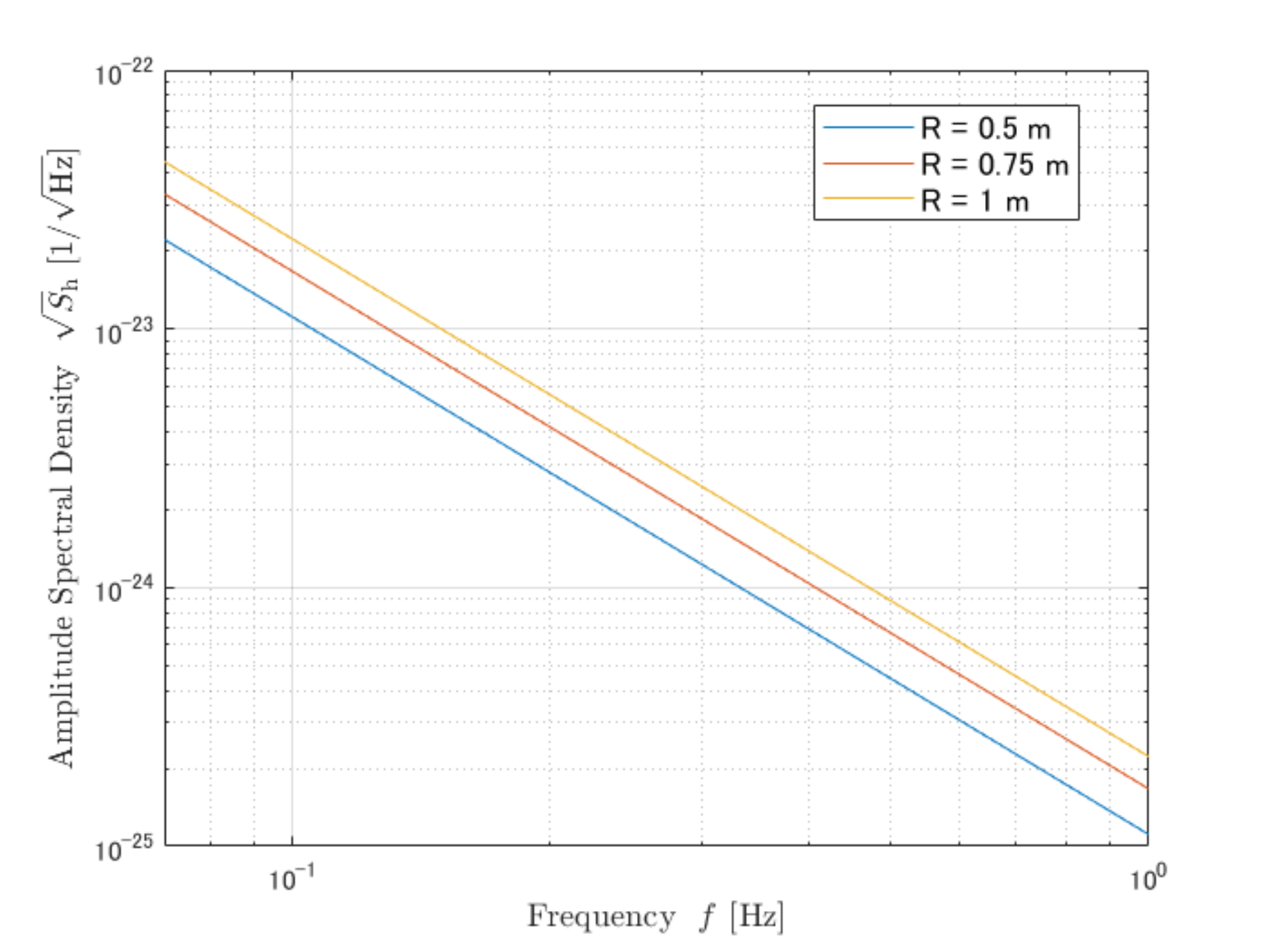}
\caption{\label{freemass}Amplitude spectral density of thermal noise in a residual gas ($10^{-8}$ Pa). The blue line shows the case of $R=0.5$ m. The red line shows the case of $R=0.75$ m. The yellow line shows the case of $R=1.0$ m.}
\end{figure}

Figure \ref{freemass} shows the amplitude spectral density of the mirror thermal noise in terms of strain $\sqrt{S_{\mathrm{h}}(f)}$, where the cavity length $L$ is 1000 km, the gas pressure is $10^{-8}$ Pa, and the mirror radius $R$ : 1 m, 0.75 m, and 0.5 m, respectively.

\vspace{-12pt}

\subsection{Internal Thermal Noise\label{int noise}}
In this subsection, to evaluate the mirror thermal noise, we use Levin's method \cite{Levin}.
At first, it is assumed that the beam radius is much smaller than the mirror radius, and the mirror is regarded as an infinite half-space.
According to Levin's method, the power spectrum is given by the following formula \cite{Levin},
\begin{equation}
\label{Levineq}
S_{x}(f)=\frac{2k_{B}T}{\pi ^2 f^2} \frac{W_{\mathrm{diss}}}{F_{0}^2}.
\end{equation}

Here, $W_{\mathrm{diss}}$ is the average power that the mirror dissipates, and $F_0$ is the peak magnitude of the pressure due to the Gaussian laser beam hitting the mirror.

Let's consider the case where the effect of friction appears in the imaginary part of Young's modulus of the mirror. 
We can calculate $W_{\mathrm{diss}}$ using two parameters, $\phi$ and  $U_{\mathrm{max}}$.
Here, $\phi$ is a parameter called the loss angle, and $U_{\mathrm{max}}$ is the elastic energy when the expansion and contraction of the mirror are maximized. 
Consequently, we obtain the power spectrum by assuming an infinite-space mirror \cite{Levin, int1, int2},
\begin{equation}
\label{int_power}
S_{x_{\mathrm{inf}}}(f)=\frac{4k_{B}T}{2\pi f}\frac{1-\sigma ^2}{\sqrt{\pi}  E_{0}r_{0}}\phi_{\mathrm{sub}}.
\end{equation} 
$\phi_{\mathrm{sub}}$ is the mechanical loss angle of the mirror. We set  $\phi_{\mathrm{sub}} = 5\times 10^{-7}$ as being representative of losses seen in fused silica mirrors.

In our simulation, the radius of the beam that hits the mirror is about the same as or larger than the mirror radius except when {$L$} is very small. 
Thus, the assumption of an infinite-space mirror does not hold, and Equation (\ref{int_power}) needs to be corrected \cite{correction}.
Specifically, it is corrected by multiplying Equation (\ref{int_power}) by a factor $C_{\mathrm{\mathrm{\mathrm{FTM}}}}$ having the laser radius $r_{0}$, the mirror radius $R$, and the mirror thickness $h$ as parameters. 
The power {spectrum} of finite-space mirror is given by,
\begin{equation}
\label{int_power2}
S_{x_{\mathrm{int}}}(f)= C_{\mathrm{FTM}}^2 \times S_{x_{\mathrm{inf}}}(f).
\end{equation} 

See Appendix \ref{appendixB} for specific $C_{\mathrm{FTM}}$ expressions.
When the mirror is infinite size, $C_{\mathrm{FTM}} = 1$.
In our calculation, we assume that the laser radius $r_0$ is the distance from the center, where the beam amplitude is $1/e$ of the maximum.  
For simplicity, we assume $r_0 = R$ at the mirror surface.
In calculating $h$, we assume that the material of the mirror is fused silica, which has a mass density of $\rho=2.196 \times 10^3 \ \mathrm{kg/m^3}$. 
We use this corrected $S_{x_{\mathrm{int}}}(f)$ in our simulation. 
\subsection{Other Sources of Thermal Noise}
In the above subsections, we show the power spectrum of the thermal noise in a residual gas (Equation (\ref{gas_power})) and the internal thermal noise (Equation (\ref{int_power})). 
In reality, the mirror has thermal noises other than the above two. Thermoelastic noise and thermal noise of optical coatings are typical examples. 
Here we demonstrate that they have a negligible effect on the calculation of SNR. 

First, assuming that the mirror has infinite size, we obtain the power spectrum of thermoelastic noise of the mirror \cite{elastic}.
\begin{equation}
\label{elas}
S_{x_{\mathrm{elas}}}(f)=\frac{16k_{B}T^2 (1+\sigma)^2 \alpha ^2 \kappa}{\sqrt{\pi} C r_{0}^3 (2\pi f)^2}.
\end{equation}

According to the formula of Equation (\ref{elas}), the effect of thermoelastic noise is small. Thermoelastic noise in a finite size mirror is larger than Equation (\ref{elas}), but still not large enough to take into consideration. 

Second, when we take the effect of optical coatings into consideration, we must multiply Equation (\ref{int_power}) with the factor $C_{\mathrm{coat}}$ \cite{Numata}; 
\begin{equation}
\label{factor}
C_{\mathrm{coat}} \sim \left( 1+ \frac{2}{\sqrt{\pi}} \frac{1-2\sigma}{1-\sigma} \frac{\phi_{\mathrm{coat}}}{\phi_{\mathrm{sub}}} \frac{d}{r_0} \right),
\end{equation}
where ${\phi_{\mathrm{coat}}}$ is mechanical loss angle of coating.
The order of the factor $C_{\mathrm{coat}}$ is determined by $d/r_{0}$. 
In this paper, to balance the diffraction loss against the expense and difficulty of large mirrors, we set the beam radius to be comparable to the mirror radius. 
Since the minimum $R$ is 1 cm in our simulation, the minimum $r_{0}$ is also 1 cm.
We assume that  $d = 4 \ \mu \mathrm{m}$ because the reflectivity of the mirror $r$ is not so high ($r\sim 0.9$).
Therefore, $d/r_{0} < 10^{-4}$ and  $C_{\mathrm{coat}} \sim 1$.

For the above reasons, we treat only thermal noise in a residual gas and internal thermal noise as noise in our simulation.

\section{Method of Optimization\label{Method of optimization}}
\begin{figure}[H]
\centering
\includegraphics[width=9 cm]{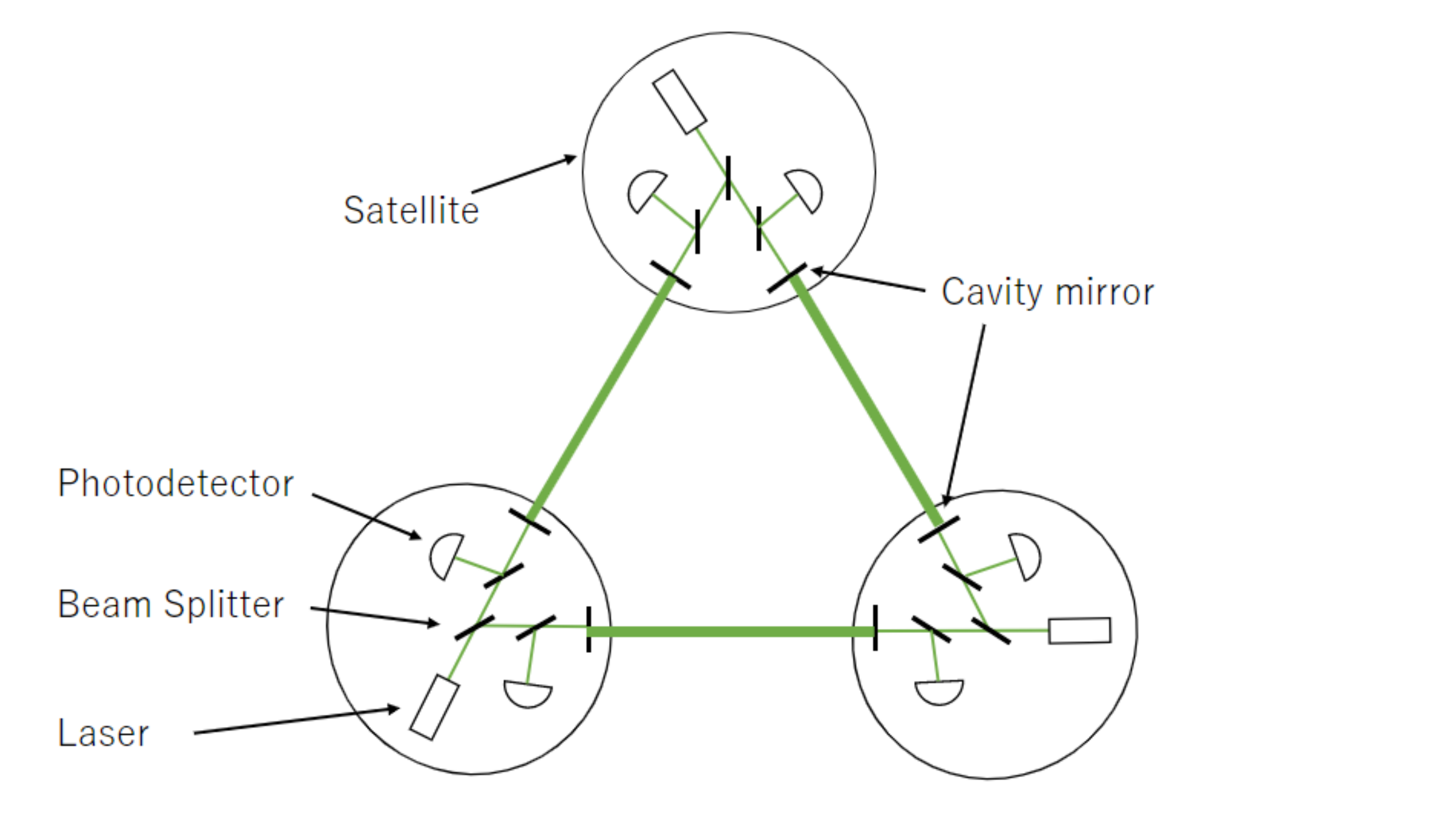}
\caption{\label{satellite} Configuration of one cluster in DECIGO. One cluster has  three satellites. Each satellite has two mirrors, and the mirrors compose FP interferometers. }
\end{figure}

DECIGO consists of four clusters that are on the heliocentric orbit of the earth. 
Two clusters are placed at the same position to detect PGW by combining them.
\mbox{Figure \ref{satellite}} shows the configuration of DECIGO's one cluster. Its features are the following:

\begin{itemize}
\item	One cluster consists of three interferometers.
\item	There are differential Fabry-Perot (FP) interferometers with 60° between each arm.
\item	Each interferometer shares each arm with two other interferometers.
\end{itemize}

Therefore, the signal and noise obtained from one DECIGO cluster must be properly considered by separating one cluster into two effective interferometers \cite{LISA}.
First, we calculate the sensitivity of one triangular cluster: $S^{\mathrm{cluster}}_{\mathrm{h}}$. Then we combine the two clusters at the same position to calculate the SNR. 
For the detection of PGW, we need two clusters because the PGW is stationary, isotropic, has no polarization, and cannot be detected by one cluster.
Then we optimize each of DECIGO’s parameters to maximize the SNR for a given mirror radius $R$. In this paper, we optimize the sensitivity of DECIGO to the PGW ($\Omega_{\mathrm{gw}} =  10^{-16} $).
The SNR for DECIGO is given by
\begin{equation}
\label{SNR1}
\mathrm{SNR}=\frac{3H_{0}^2}{10\pi ^2}\sqrt{T_{\mathrm{obs}}} \left[\int_{f_{\mathrm{min}}}^{1} \frac{2\gamma ^{2} (f) \Omega_{\mathrm{gw}}^2 (f)}{f^6 P_1 (f)P_2 (f)} df\right]^{\frac{1}{2}},
\end{equation}
where $P_1(f) = P_2(f) = S^{\mathrm{cluster}}_{\mathrm{h}}/5$, $T_{\mathrm{obs}}$ = 3 years. $f_{\mathrm{min}}$ is calculated for the two cases of 0.1~Hz and 0.07 Hz. 
$\gamma$ is the normalized overlap reduction function, equivalent to 1 because the two clusters are at the same position.

Considering quantum noises: shot noise $S_{\mathrm{h}_{\mathrm{shot}}}(f)$ and radiation pressure noise $S_{\mathrm{h}_{\mathrm{rad}}}(f)$ in addition to thermal noise, $S^{\mathrm{cluster}}_{\mathrm{h}}$ is given by  \cite{Ishikawa},

\begin{equation}
\label{cluster2}
S^{\mathrm{cluster}}_{\mathrm{h}}(f)=\frac{5\sqrt{2}}{3\mathrm{sin}^{2}\frac{\pi}{3}}\left[(\sqrt{S_{\mathrm{h}_{\mathrm{shot}}}(f)})^{2}+(\sqrt{S_{\mathrm{h}_{\mathrm{rad}}}(f)})^{2}+(\sqrt{S_{\mathrm{h}_{\mathrm{gas}}}(f)})^2+(\sqrt{S_{\mathrm{h}_{\mathrm{int}}}(f)})^2\right],
\end{equation}

where 
\begin{align}
\label{shot}
\sqrt{S_{\mathrm{h}_{\mathrm{shot}}}}(f)&=\frac{1}{4\pi L}\frac{(1-r_{\mathrm{eff}}^2)^2}{t_{\mathrm{eff}}(tD)r_{\mathrm{eff}}} \sqrt{\frac{4\pi \hbar c \lambda}{P_{0}}}\sqrt{1+(\frac{f}{f_{\mathrm{p}}})^2},\\
\label{rad}
\sqrt{S_{\mathrm{h}_{\mathrm{rad}}}}(f)&=\frac{4}{mL(2\pi f)^2}\frac{t_{\mathrm{eff}}^2(rD)^2(1+r_{\mathrm{eff}}^2)}{(1-r_{\mathrm{eff}}^2)^2}\sqrt{\frac{\pi \hbar P_0}{c\lambda}}\sqrt{\frac{1}{1+(\frac{f}{f_\mathrm{p}})^2}},
\end{align}
and  $\sqrt{S_{\mathrm{h}_{\mathrm{gas, int}}}} =  \sqrt{S_{x_{\mathrm{gas, int}}}}/L$.

In Equation (\ref{cluster2}), the coefficient $\frac{5\sqrt{2}}{3\mathrm{sin}^{2}\frac{\pi}{3}}$ represents that  one DECIGO cluster consists of three interferometers with $\pi/3$ arm angle.

Table \ref{tab2} shows the meaning of each symbol in Equations (\ref{SNR1})--(\ref{rad}).
\begin{table}[H] 
\centering
\caption{Meaning of each symbol.\label{tab2}}
\setlength{\tabcolsep}{7.5mm}\begin{tabular}{cc}
\toprule
\textbf{Symbol}	& \textbf{Meaning}	\\
\midrule
$L$		& Cavity length\\
$m$		& Mirror mass	\\
$P_{0}$      &Laser power entering beam splitter\\
$\lambda$  (=515 $\times$ $10^{-9}\ \mathrm{m})$ &Laser wavelength\\
$r$ &Mirror reflectivity\\
$t$ & Mirror transmissivity\\
$D$ & Effect of diffraction loss\\
$r_{\mathrm{eff}} \equiv rD^2$& Effective mirror reflectivity\\
$t_{\mathrm{eff}} \equiv tD^2$ & Effective mirror transmissivity\\
$c$ (=2.9979 $\times$ $10^8\  \mathrm{m/s})$ & Light speed\\
$\hbar$ (=1.0546 $\times$ $10^{-34}\ \mathrm{J s} )$ & Planck constant\\
$H_{0}$ (=70/3.086 $\times$ $10^{19} \ \mathrm{km/s/Mpc})$ &Hubble constant\\
$\mathcal{F} \equiv \pi r/(1-r^2)$ &Finesse\\
$f_p \equiv c/4F_{\mathrm{eff}}L $ & Cavity pole frequency\\
$\mathcal{F}_{\mathrm{eff}} \equiv \pi r_{eff}/(1-r_{\mathrm{eff}}^2)$ & Effective finesse\\

\bottomrule
\end{tabular}
\end{table}

In the following section, we use the maximized $D = D_{\mathrm{max}}$,
\begin{equation}
D^2_{\mathrm{max}}=1-\mathrm{exp}\left[ -\frac{2\pi}{L\lambda}R^2\right]. 
\end{equation}

The cavity setting for the maximizing $D$ is shown in Appendix \ref{appendixC}.   
We optimize each of DECIGO’s parameter to maximize the SNR for a given mirror radius $R$. 
\subsection{Treatment of Each Noise in the Simulation}
In this subsection, we show how to treat each noise in our simulation. 

For GWs from DWDs, we consider two patterns of the cutoff frequency: 0.07 Hz and 0.1 Hz.  
Considering the power spectrum of all the noises, we calculate the noise power spectrum of one cluster $S^{\mathrm{cluster}}_{\mathrm{h}}(f)$ by taking the sum of squares with shot noise and radiation pressure noise because all the noises are independent (see Equation (\ref{cluster2})). 

Note that, $\sqrt{S_{\mathrm{h}_{\mathrm{shot}}}}$ and $\sqrt{S_{\mathrm{h}_{\mathrm{rad}}}}$ are the strain of quantum noise of one interferometer that has a $90^\circ$ arm angle.
The strain of thermal noise of one mirror is $\sqrt{S_{\mathrm{h}_{\mathrm{gas, int}}}}$.  
Since one arm has two mirrors and their thermal noises are independent of each other, the strain of one arm is $\sqrt{\mathrm{2}}$ times larger than that of one mirror. When two arms of one interferometer are correlated, the noise is $\sqrt{\mathrm{2}}$ times larger than that of one arm, and the signal is 2 times larger. 
Therefore, the strain of thermal noise of one interferometer is represented as  $\sqrt{S_{\mathrm{h}_{\mathrm{gas, int}}}}$.

\subsection{Method of Calculation}
We calculate the SNR of DECIGO, applying Equation (\ref{cluster2}) as a function of $R, L, r$, and~$P_{0}$. 
\begin{equation}
\mathrm{SNR} = \mathrm{SNR} (R, L, r, P_0).
\end{equation}

Further, we decide the optimized $L,$ $r$, and $P_{0}$ that give the maximum SNR for a given $R$.

In our simulation, we consider two cases of gas pressure in the satellite:  $10^{-8}$ Pa (high-density-gas case) and $10^{-9}$ Pa (low-density-gas case).
Considering these two cases, we estimate the magnitude of the internal pressure required for the DECIGO's satellites. 
 
In addition, we consider two patterns for the mirror: the constant-mirror-mass model and the constant-mirror-thickness model. 
In the constant-mirror-mass model, we set the mirror mass to be 100 kg regardless of $R$.
In the constant-mirror-thickness model, we set the mirror mass to be proportional to the square of $R$. 

\begin{equation}
m = \left(\frac{R}{0.5\ \mathrm{m}}\right) ^2 \times 100\ \mathrm{kg}.
\end{equation}
$m$ is decided to be 100 kg at $R=0.5\ \mathrm{m}$, which is the default value of DECIGO.
{In this paper, we consider two mirror models as frameworks with which we can make a further optimization after the limitations of the mirror mass and size are set.}
We calculate the SNR over the limited range of each parameter shown in Table \ref{tab3}. 
In addition, we show the results of the optimizations in each case shown in Table \ref{tab4}.
That is, we show eight results obtained by combining each model.
\begin{table}[H] 
\centering
\caption{Limited range of each parameter.\label{tab3}}
\setlength{\tabcolsep}{27.2mm}\begin{tabular}{cc}
\toprule
\textbf{Symbol}	& \textbf{Range}	\\
\midrule
$R$ &0 to 1 m\\
$r$	&0 to 1\\
$P_0$& 0 to 100 W\\
$L$ & No limit\\
\bottomrule
\end{tabular}
\end{table}

\vspace{-9pt}
\begin{table}[H] 
\centering
\caption{DECIGO's parameters that have different values depending on the model.\label{tab4}}
\setlength{\tabcolsep}{5.8mm}\begin{tabular}{cc}
\toprule
\textbf{Parameter}	& \textbf{Value in Each Model}	\\
\midrule
Cutoff frequency &0.07 Hz/0.1 Hz\\
Pressure in the satellite &$10^{-8}$ Pa/$10^{-9}$ Pa\\
Mirror mass &Constant (100 kg)/Proportional to the square of R.\\
\bottomrule
\end{tabular}
\end{table}
\section{Result\label{Results}}

\subsection{Optimization of SNR and Parameters}
\begin{figure}[H]

\centering 
 \begin{tabular}{cc}

  \begin{minipage}[]{7cm}
    \centering
    \includegraphics[width=7cm]{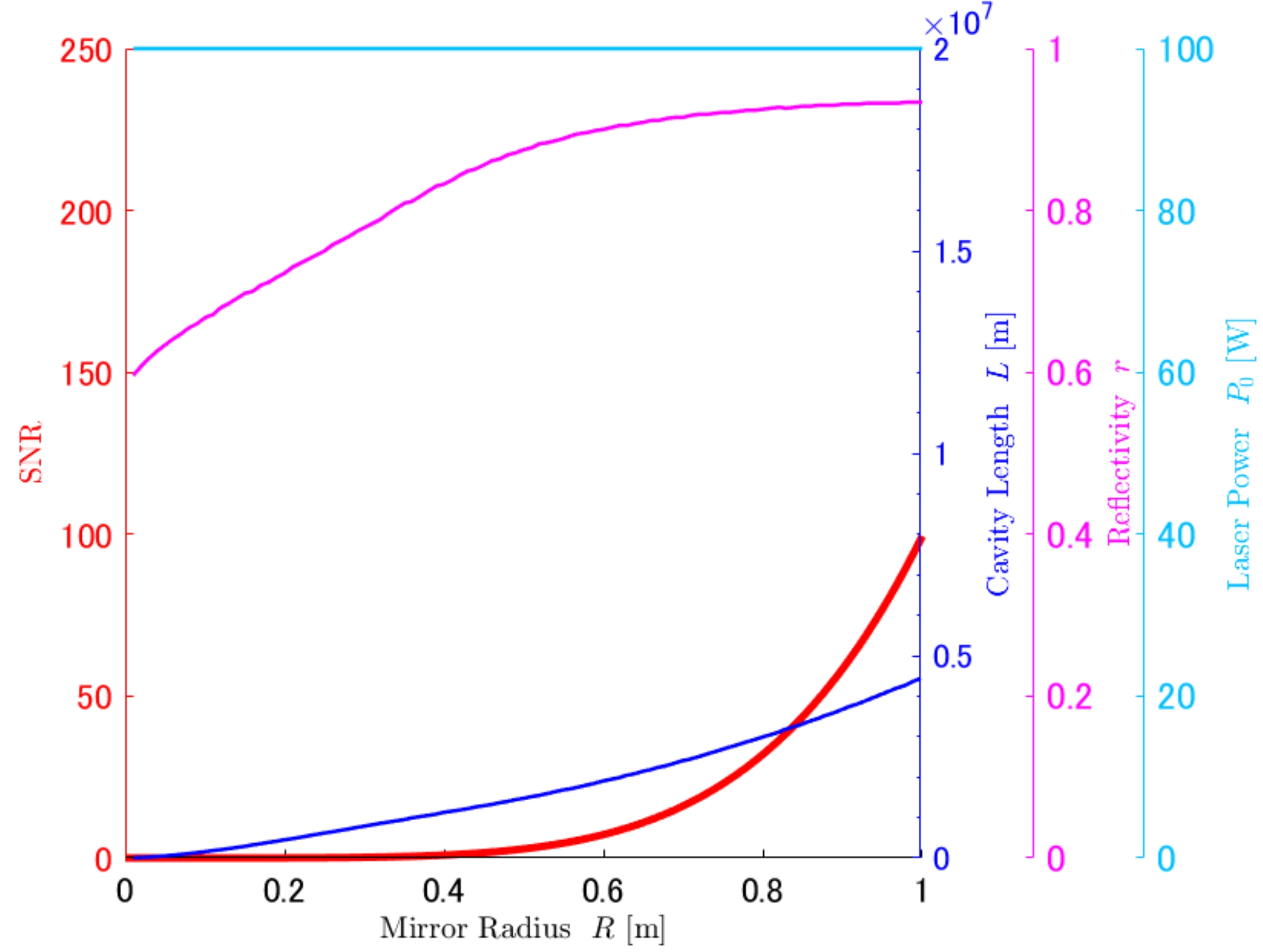}
    \subcaption{Constant-thickness model in the high-density-gas case
}
   \end{minipage}

  \begin{minipage}[]{7cm}
    \centering
    \includegraphics[width=7cm]{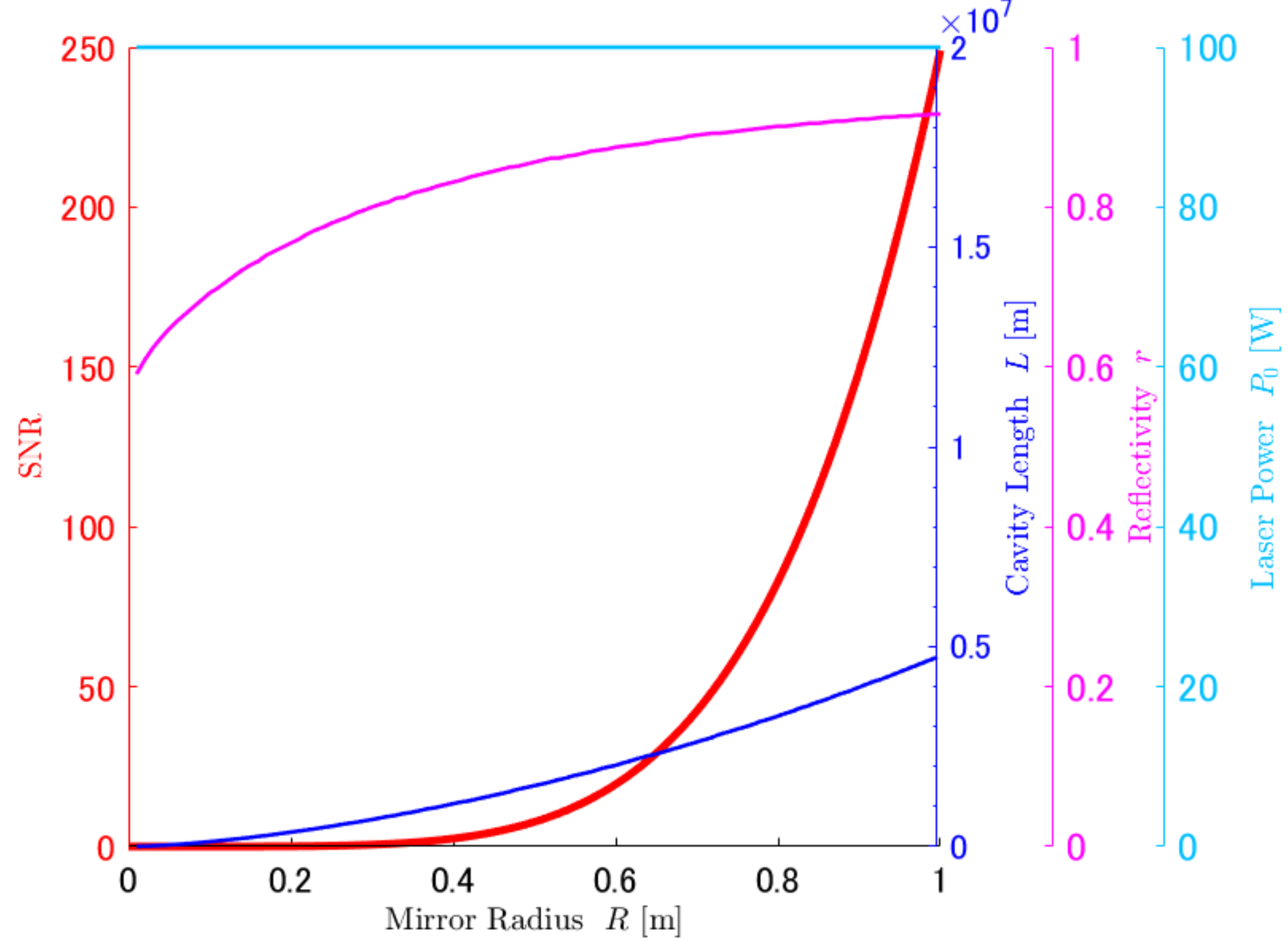}
    \subcaption{Constant-thickness model in the low-density-gas case
}
  \end{minipage}
\\
\\
 \begin{minipage}[]{7cm}
    \centering
    \includegraphics[width=7cm]{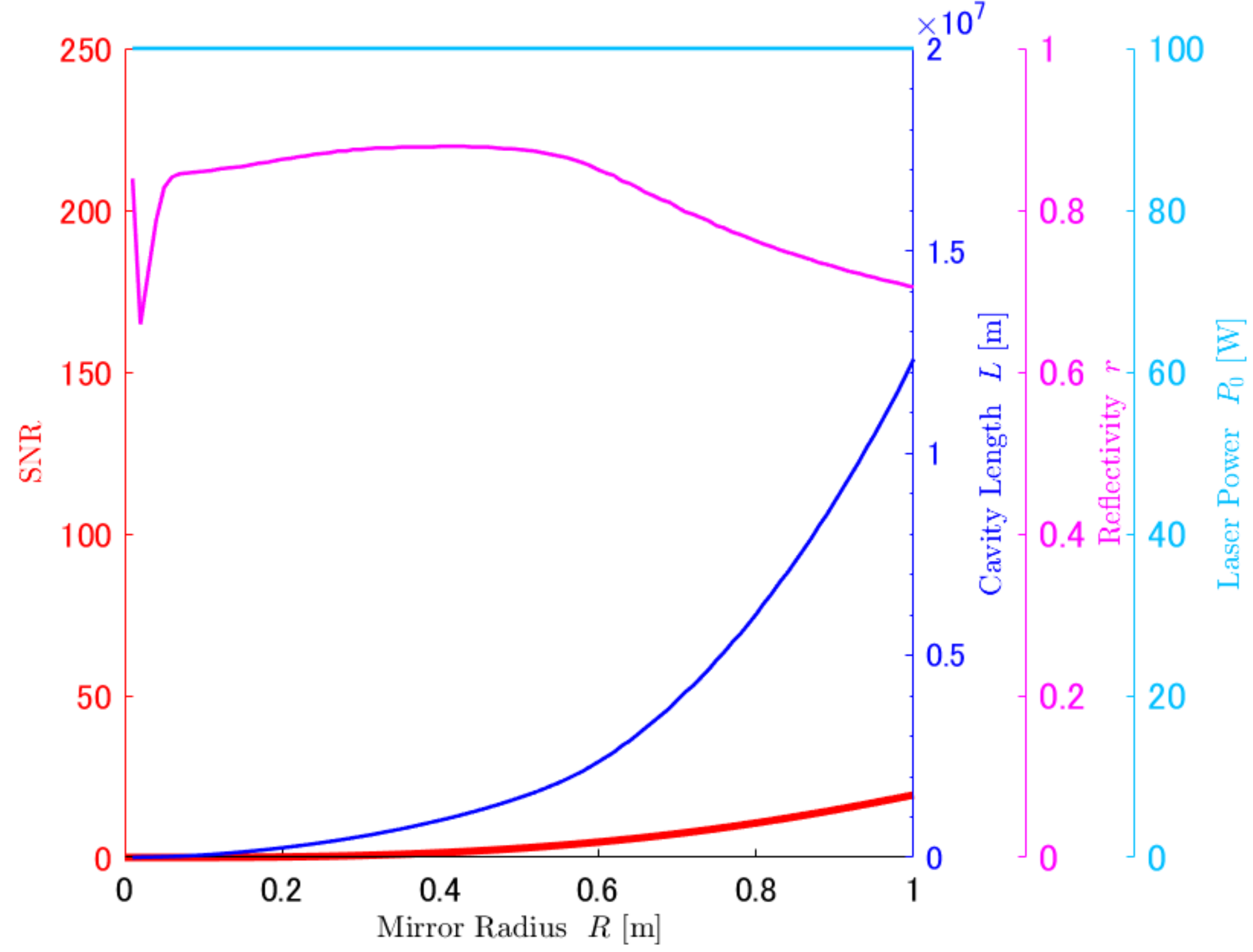}
    \subcaption{Constant-mass model in the high-density-gas case.
}
   \end{minipage}

 \begin{minipage}[]{7cm}
    \centering
    \includegraphics[width=7cm]{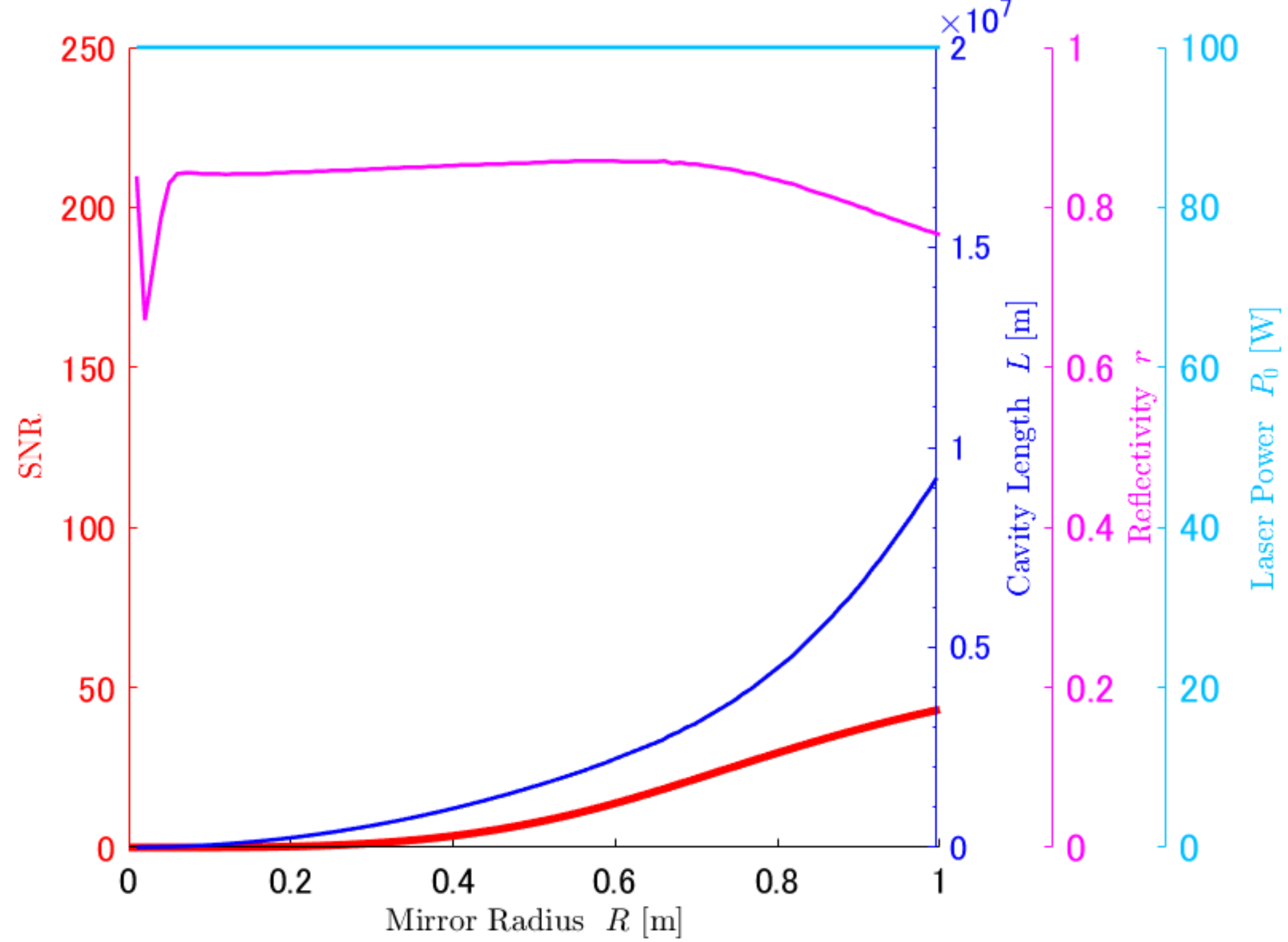}
    \subcaption{Constant-mass model in the low-density-gas case.
}
  \end{minipage}
\\
\end{tabular}

\caption{\label{results1}Maximized SNR for
$R$ (red line) and optimized $L$ (blue line), $r$ (magenta line), and $P_{0}$ (cyan line) in the optimistic DWD model. (\textbf{a},\textbf{b}) show the results of the constant-thickness model. (\textbf{c},\textbf{d}) show the results of the constant-mass model.
 (\textbf{a},\textbf{c}) show the results of the high-density-gas case. (\textbf{b},\textbf{d}) show the results of the low-density-gas case.} 
\end{figure}

\begin{figure}[H]

\centering 
 \begin{tabular}{cc}
  \begin{minipage}[]{7cm}
    \centering
    \includegraphics[width=7cm]{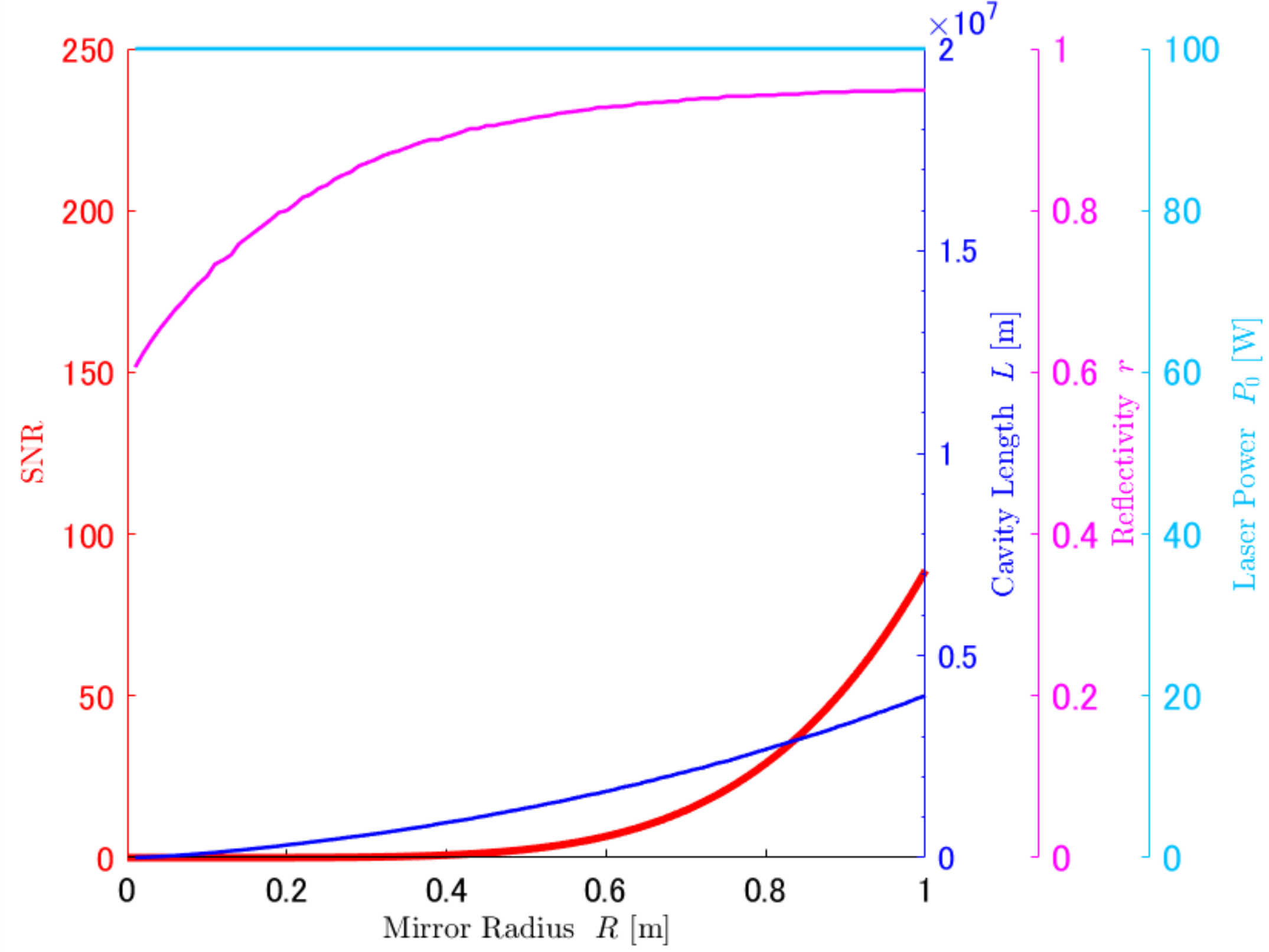}
    \subcaption{Constant-thickness model in the high-density-gas case.
}
  \end{minipage}

 \begin{minipage}[]{7cm}
    \centering
    \includegraphics[width=7cm]{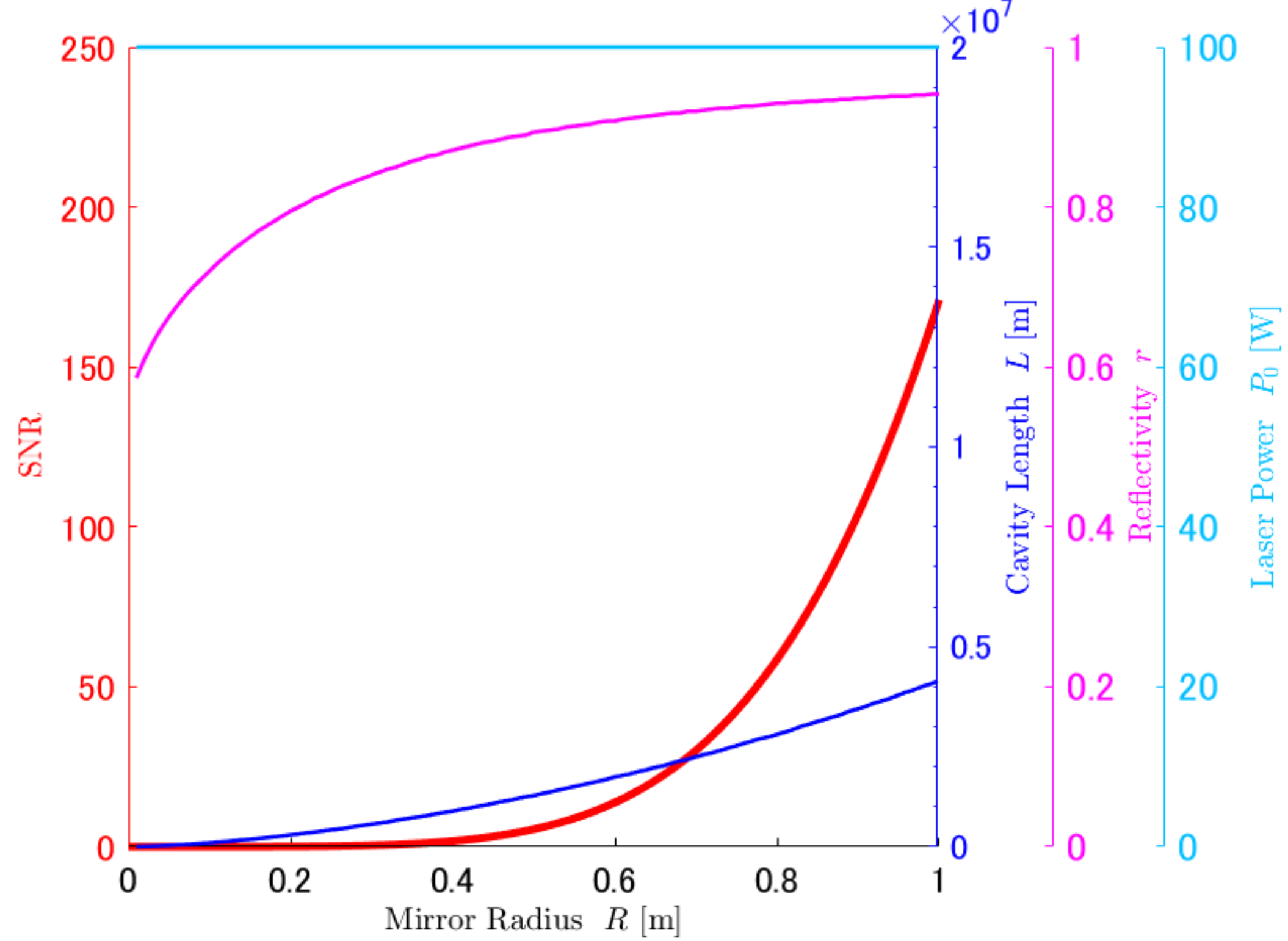} 
    \subcaption{Constant-thickness model in the low-density-gas case.
}
  \end{minipage}
\\
\\
 \begin{minipage}[]{7cm}
    \centering
    \includegraphics[width=7cm]{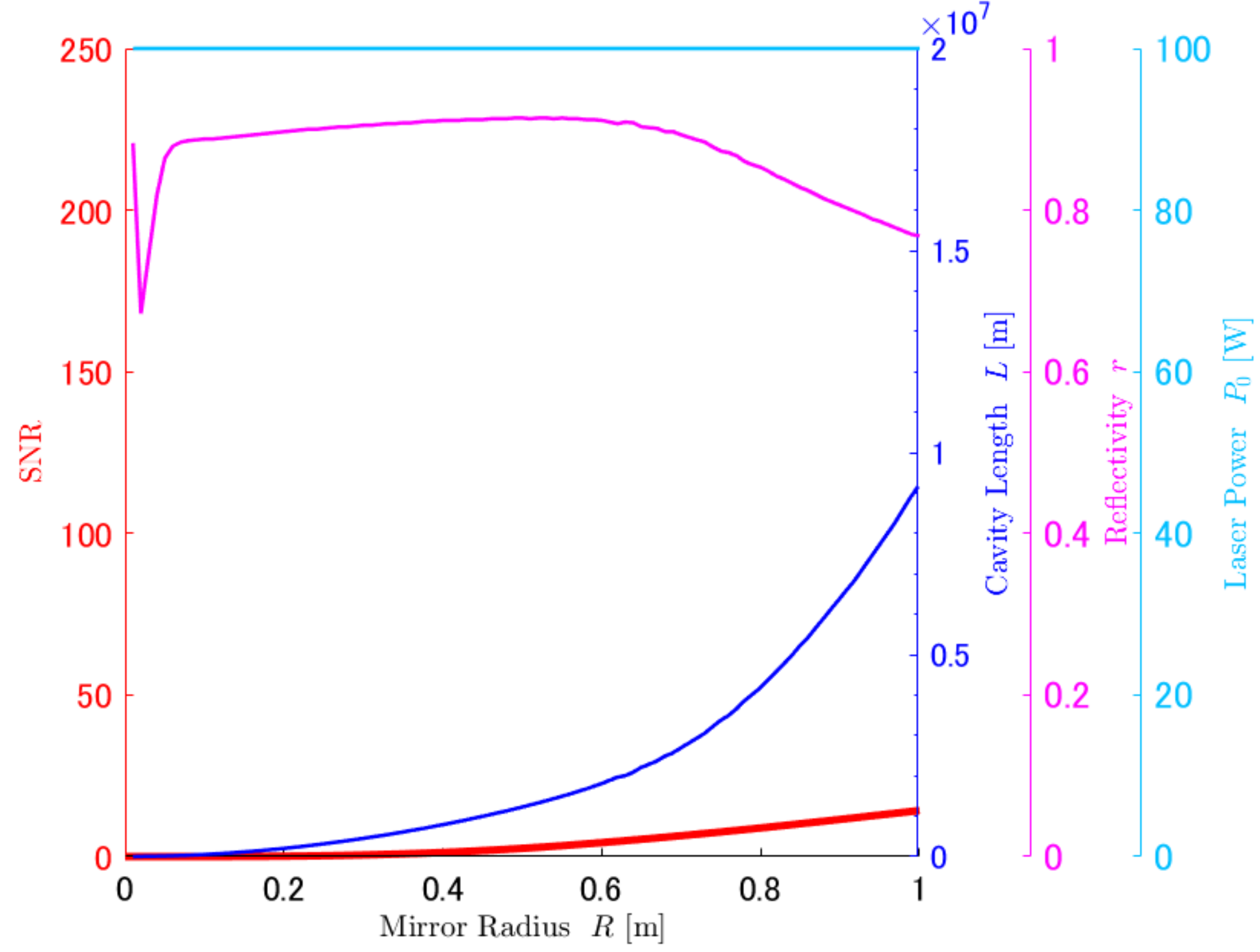}
    \subcaption{Constant-mass model in the high-density-gas case.
}
  \end{minipage}

  \begin{minipage}[]{7cm}
    \centering
    \includegraphics[width=7cm]{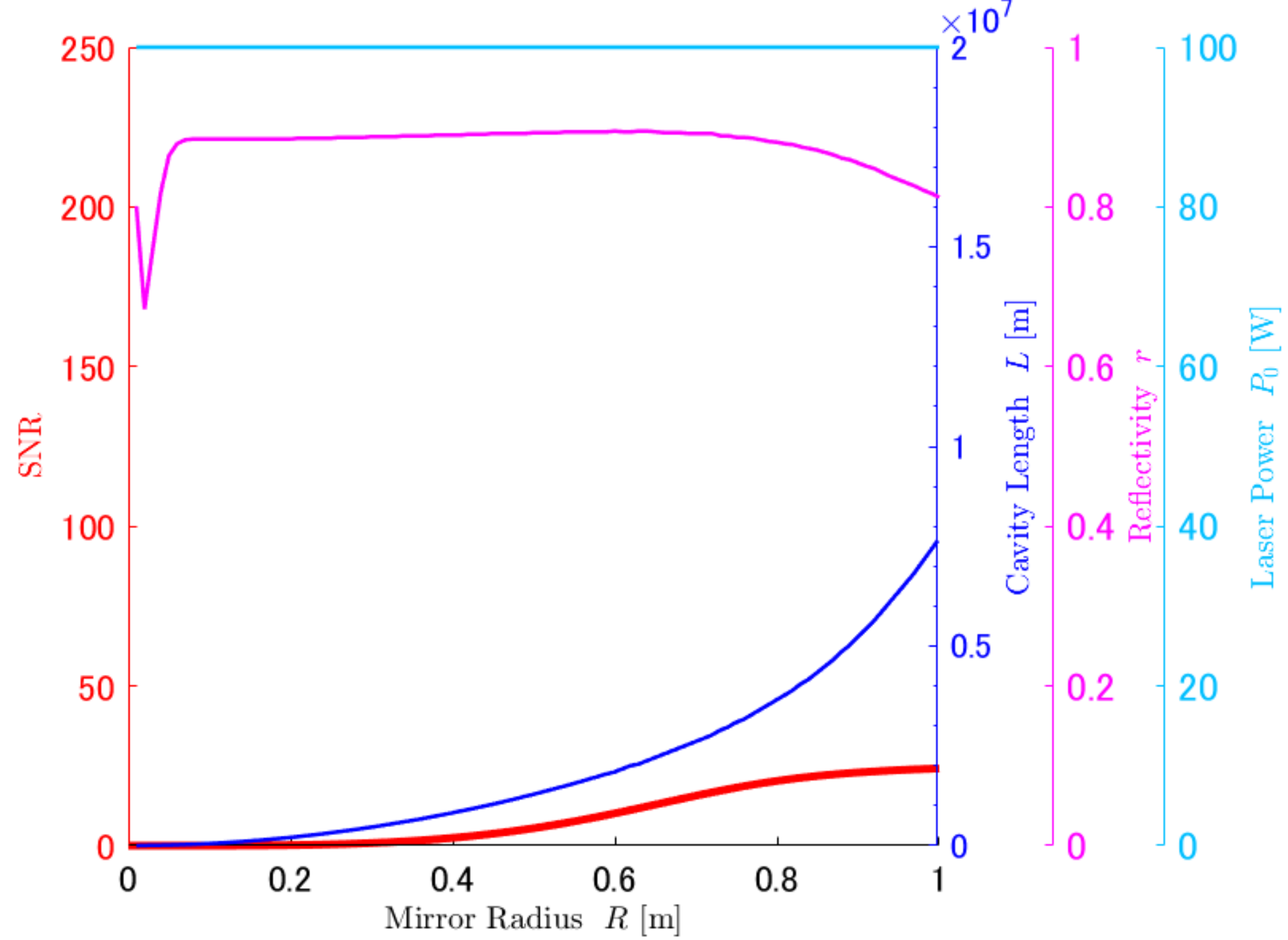}
    \subcaption{Constant-mass model in the low-density-gas case.
}
  \end{minipage}
\\
\end{tabular}
\caption{\label{results2}Maximized SNR for $R$ (red line) and optimized $L$ (blue line), $r$ (magenta line), and $P_{0}$ (cyan line) in the standard DWD model. (\textbf{a},\textbf{b}) show the results of the constant-thickness model. (\textbf{c},\textbf{d}) show the results of the constant-mass model. (\textbf{a},\textbf{c}) show the results of the high-density-gas case. (\textbf{b},\textbf{d}) show the results of the low-density-gas case.}

\end{figure}

In this section, we show the result of the optimization in Figure \ref{results1} (optimistic DWD model) and Figure \ref{results2}  (standard DWD model). 

Figures \ref{results1} and \ref{results2} show the maximized SNR and optimized parameters $L$, $r$, and $P_{0}$ as a function of $R$.
In Figures \ref{results1} and \ref{results2}a,b show the results of the constant-mirror-thickness model. 
Figure
\ref{results1} and \ref{results2}c,d show the results of the constant-mirror-mass model. 
Figure \ref{results1} and \ref{results2}a,c show the results of the high-density-gas case.
Figure \ref{results1} and \ref{results2}b,d show the results of the low-density-gas~case.

First, the figures show very similar characteristics in both DWD models.

In all figures of Figures \ref{results1} and \ref{results2}, the optimized $P_{0}$ is 100 W. The maximized SNR and the optimized $L$ increase as the mirror radius $R$ increases. This is because the noise strain is scaled by 1/$L$.
However, the optimized $r$ has different characteristics in the constant-thickness model and constant-mass model.
The optimized $r$ increases as the mirror radius $R$ increases in the constant-thickness model. 
On the other hand, the optimized $r$ has two characteristics in the constant-mass model. 
The first one is that it has a dip when $R$ is small due to the extreme cylindrical shape of the mirror, which increases internal thermal noise. 
The other one is that it decreases with the increase of $R$ because the dominant noise source depends on $R$.

The features of optimized parameters mentioned in the above paragraph can be explained by the characteristics of shot noise and thermal noises.
The large $R$ increases thermal noises and decreases the effect of diffraction loss. 
At the same $R$, the large $P_{0}$ and the finesse of cavity decreases the shot noise, and the long $L$ decreases the strain of thermal noise (see Equation (\ref{cluster2})).
When $L$ is extremely long, the effect of diffraction loss is large.
If the finesse $F$ (see Table \ref{tab2}) is too high in a situation where the diffraction loss is large, the effect of losing the laser power due to the diffraction loss is greater than the effect of amplifying the laser power in the cavity.
Thus, the laser power that can be detected decreases, and the shot noise increases.
To reduce the shot noise, it is necessary to lower $r$ to reduce the finesse to some extent. 
According to the characteristics of shot noise and thermal noises, when $R$ is small, thermal noises do not matter. Thus, shot noise is dominant when $R$ is small. On the other hand, when $R$ is large, thermal noises are dominant.
Since the magnitude of each noise has a continuous dependence on $R$, the dominant noise is swapped between the shot noise and thermal noises at a specific $R$.
In the range of $R$, where thermal noises are dominant, the optimized $L$ is long in order
constant-mirror-thickness model at the same $R$. This is because the internal thermal noise due to the distortion of the mirror is significantly larger when $R$ is large. 

Let us focus on each gas case. The maximized SNR in the low-density-gas case is about 2 times larger than that in the high-density-gas case. This result indicates that the thermal noise in a residual gas has a significant impact on the total noise spectrum in this model.

\begin{figure}[H]
\centering
\includegraphics[width=10cm]{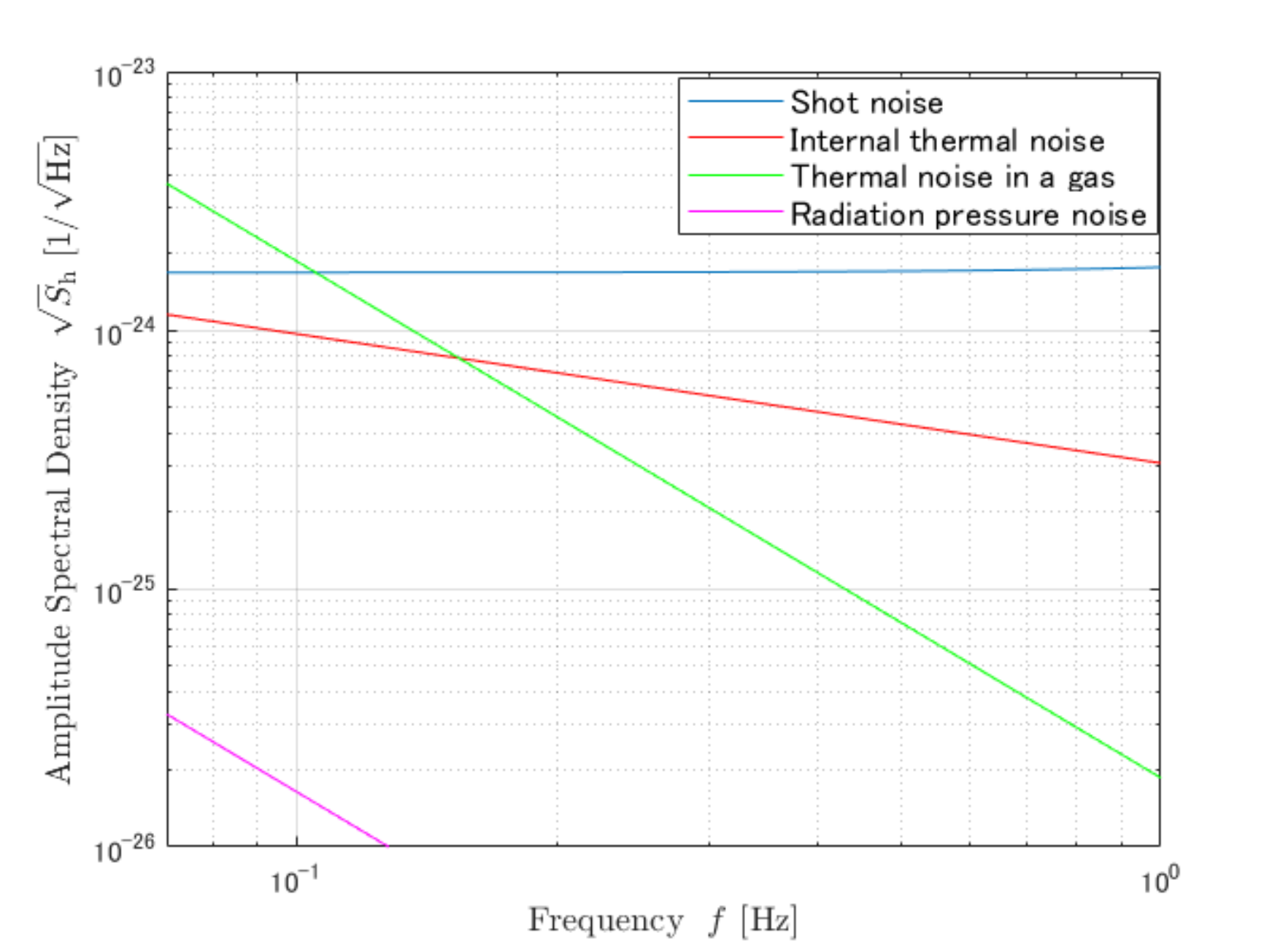}
\caption{\label{shot_thermalnoise}Shot noise (blue line), thermal noise in a residual gas (green line), internal thermal noise (red line), and radiation pressure noise (magenta line) of Figure \ref{results1}c.
All of them are optimized with $L=1.2\times 10^7\ \mathrm{m}$, $r=0.75$ at $R=1\ \mathrm{m}$.} 

\end{figure}

It is impossible to reduce all noises at the same time by changing $R$, $L$, and $r$.
The SNR is maximized when the dominant noises have approximately the same magnitude in the lower frequency band.
Actually, at $R$ = 1 m, the shot noise, thermal noise in the high-density-gas case, and internal thermal noise of the constant-mirror-mass model with optimized parameters are about the same around 0.1 Hz (see Figure \ref{shot_thermalnoise}).

\subsection{Comparison of the Estimated Strain Sensitivities, Especially at Large \textit{R}}
\begin{figure}[H]
\centering 
\begin{tabular}{cc}
 \begin{minipage}[]{7cm}
    \includegraphics[width=\textwidth]{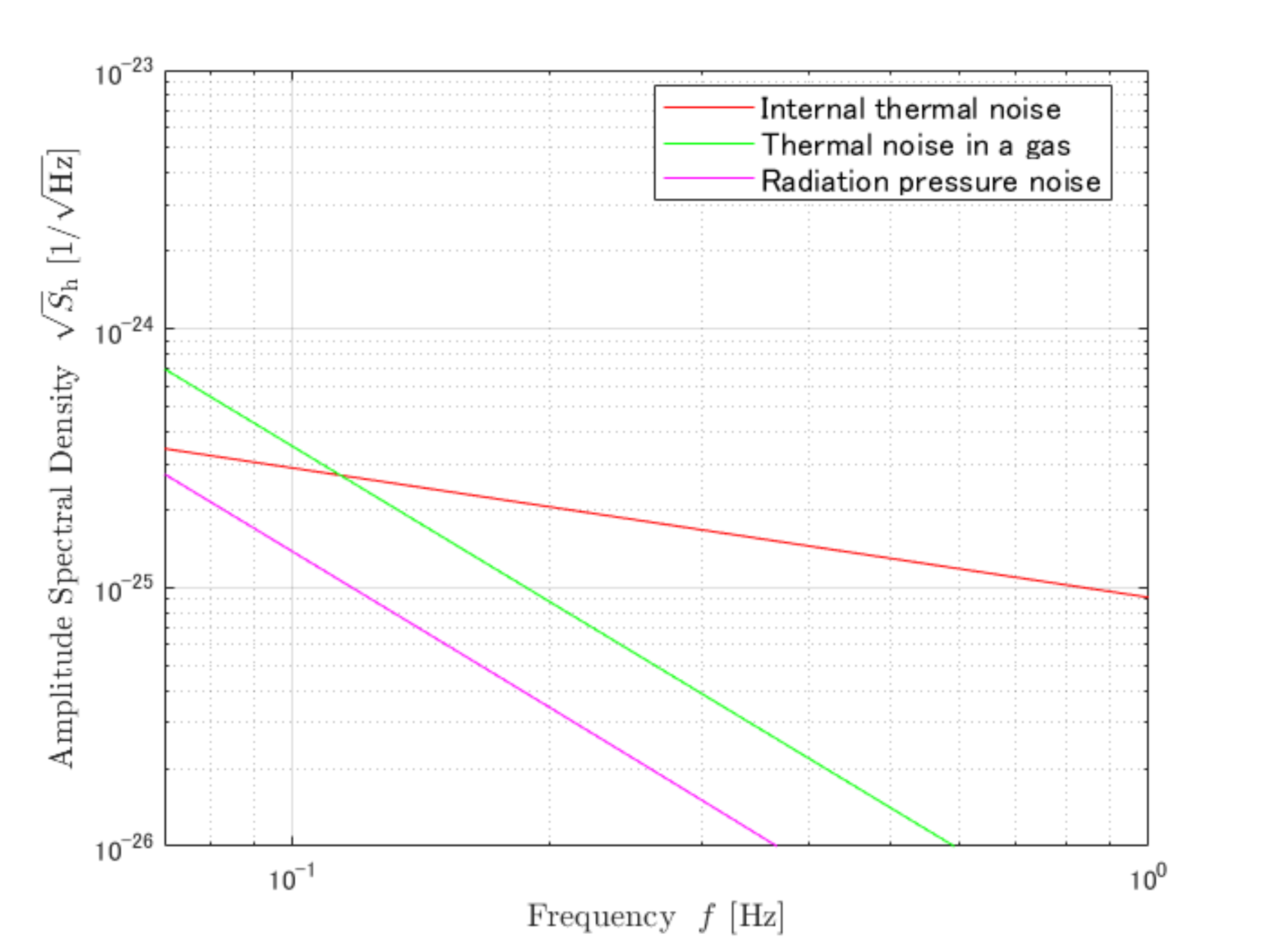}
    \subcaption{Constant-mirror-thickness model: strain-equivalent noise due to internal thermal noise (red line), thermal noise in a low-density-gas case (green line), and radiation pressure noise (magenta line) of the constant-mirror-thickness model }
  \end{minipage}
\hfill
  \begin{minipage}[]{7cm}
 \centering
    \includegraphics[width=\textwidth]{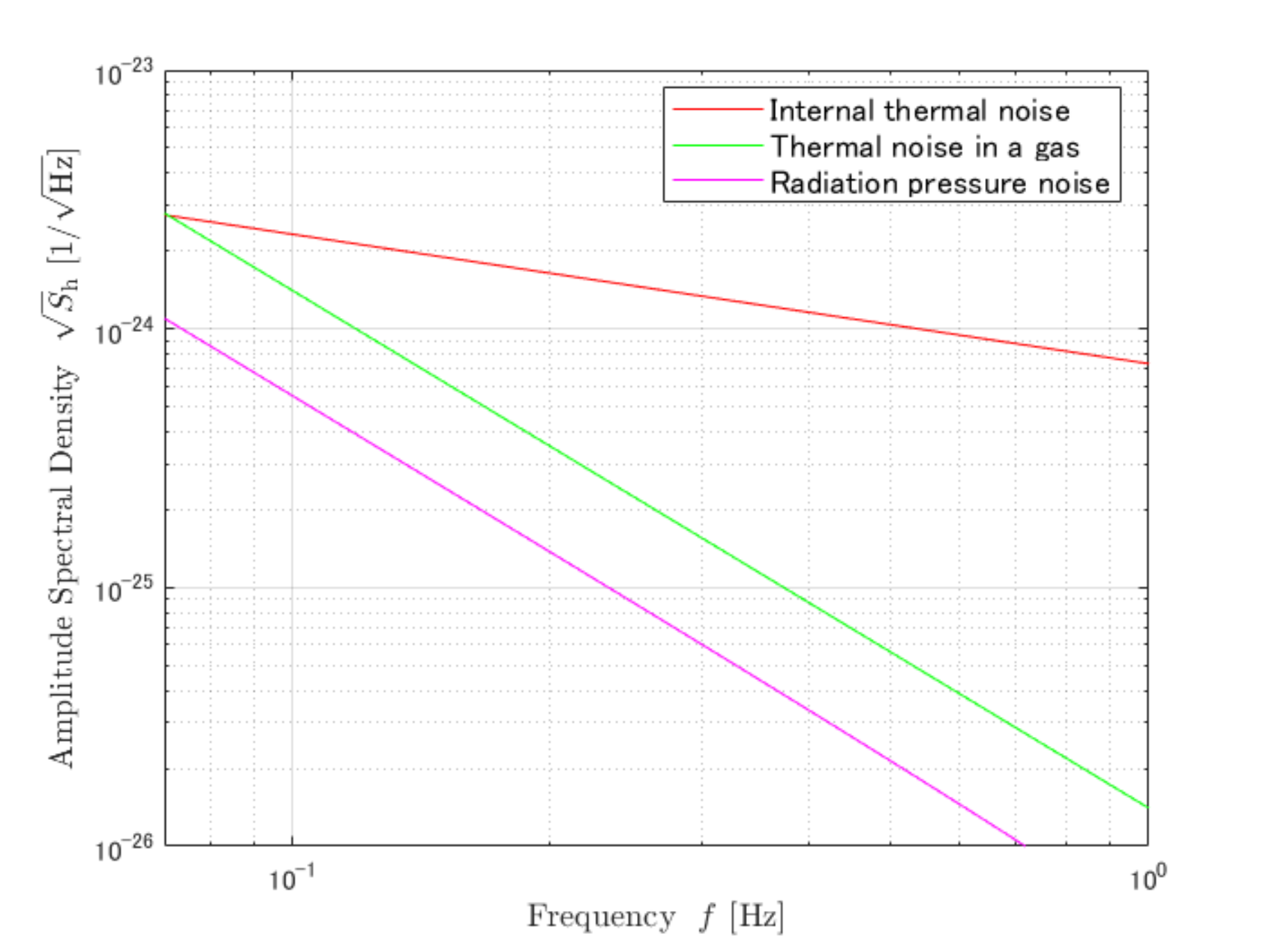}
    \subcaption{Constant-mirror-{mass} model: strain-equivalent noise due to internal thermal noise (red line), thermal noise in a low-density-gas case (green line), and radiation pressure noise (magenta line) of the constant-mirror-mass model.}
  \end{minipage}
  \\
  \\
\end{tabular}
\caption{\label{results5}Strain of thermal noises and radiation pressure noise at $R=1\ \mathrm{m}$, $L=5\times 10^6\ \mathrm{m}$, $r=0.9$, and $P_{0}=100\ \mathrm{W}$.
(\textbf{a}) shows the optimized noises.  (\textbf{b}) shows the non-optimized noises. }
\end{figure}

The purpose of this paper is to consider the parameters that give a larger SNR; thus, we consider the noise when $R$ = 1 m, which has the highest SNR in all cases.
In this subsection, we compare the difference in the noise due to the difference in the mirror shape.
We note that shot noise does not depend on the mirror thickness and mass.
Figure \ref{results5} shows the strains of the two thermal noises in the low-density-gas case and radiation pressure noise.
{In order to show only the effect of mirror models, we set each parameter (excluding mirror thickness) to the optimized value in Figure \ref{results1}b.}

In Figure \ref{results5}, since the magnitude of the radiation pressure noise is one-third of the thermal noises, we focus on the relationship between the thermal noises and the shape of the mirror.
Figure \ref{results5} shows that thermal noise due to the gas is 4 times different because the mirror mass and volume are 4 times greater for the constant-mirror-thickness model than for the constant-mirror-mass model. 
The internal thermal noise has a difference of a magnitude of 10 times between the two mirror models.

The difference in the internal noise between the two mirror models can be explained by the difference in the mirror shape.
In the constant-mirror-mass model, which is equal to the constant-mirror-volume model, the mirror thickness $h$ is proportional to $R^{-2}$.
On the other hand, in the constant-mirror-thickness model, $h$ does not depend on $R$.
The dependency of $r_0/h$ differs between the two models by $R^2$.
The factor $C_{\mathrm{FTM}}$ for considering the size of the mirror introduced in Section \ref{int noise} is determined by the ratio of the laser radius to the thickness of the mirror $r_{0}/h$.
{When $R$ is large,} the increase of $r_{0}/h$ increases $C_{\mathrm{FTM}}$ {(see Figure \ref{CFTM})}.
Therefore, the internal thermal noise of  the constant-mirror-mass model is larger than that of  the constant-mirror-thickness model.

\section{Conclusions\label{Conclusion}}
We obtained the optimum parameters that maximize SNR of two correlated DECIGO detector clusters with gravitational waves from double white-dwarf binary systems and detector thermal noises in addition to the quantum noise including the effect of diffraction loss. 
We have found that we can obtain an extremely good SNR from the most 
{optimistic} model among all models we treated in this paper (Figure \ref{results1}b). 
In addition, we have also found that the characteristics of the optimized DECIGO's parameters $L$, $r$, and $P_0$ are independent of the DWD's cutoff frequency. 
Focusing on the DECIGO design, we have found that making the mirror heavier could reduce the total noise.
The mirror mass for the best SNR in this paper is four times as large as the default value. 
For future work, it is necessary to consider the DECIGO's mirror and its surroundings in more detail to improve the accuracy of the simulation. 
In addition, it is also necessary to investigate the feasibility of large and heavy mirrors to determine the DECIGO parameters for the improvement of the detectability of the PGW.
{The limitation of mirror mass and size will be determined by the launch capacity of a satellite and the progress in technological development. In the future, we will investigate the limitations of the mirror mass and size to determine the optimum design for DECIGO.}

\section*{Acknowledgments}
We would like to thank Kenji Numata and Kentaro Komori for helpful discussion about thermal noises, Tomoya Kinugawa and Gijs Nelemans for helpful advice about gravitational waves from white dwarf binaries, Kazuhiro Nakazawa for helpful advice about the environment in the satellite, and David H. Shoemaker for the editorial comments.

\begin{appendices}

\section{Derivation of Factor \textit{b}}
\label{appendixA}
Here, we derive Equations (\ref{gas_power}) and (\ref{gas_factor}).
Under the model of the mirror and its surroundings in Section \ref{gasgas},
when the mirror is stationary, gas molecules hit the mirror. 
On average, the same number of molecules collide on both sides of the mirror. 
The number of molecules of average velocity $\bar{v}$ coming from one side per unit time is 

\begin{equation}
N=\frac{1}{4}n\bar{v}S.
\end{equation}

Then, we consider the frictional force $F_{\mathrm{fric}}$ that the mirror with velocity $v_{p}$ receives
\begin{equation}
F_{\mathrm{fric}}=-\frac{1}{4}nS\mu \bar{v} v_{p} =-bv_{p}.
\end{equation}

The equation of motion of this system is written as  
\begin{equation}
F_{\mathrm{ext}}=m\ddot{x}+2b\dot{x}.
\end{equation}

The admittance $Y(f)$ is
\begin{equation}
\label{Y}
Y=\frac{1}{i2\pi fm+b}.
\end{equation}

Inserting Equation (\ref{Y}) into Equation (\ref{FDT}), we obtain the power spectrum $S_{x}(f)$ of \mbox{Equation (\ref{gas_power})}.
In this paper, assuming that the gas is nitrogen, we calculate $b$ of Equation (\ref{gas_factor}).
\section{Effect of Diffraction Loss}
\label{appendixC}
\begin{figure}[H]
\centering
\includegraphics[width=11 cm]{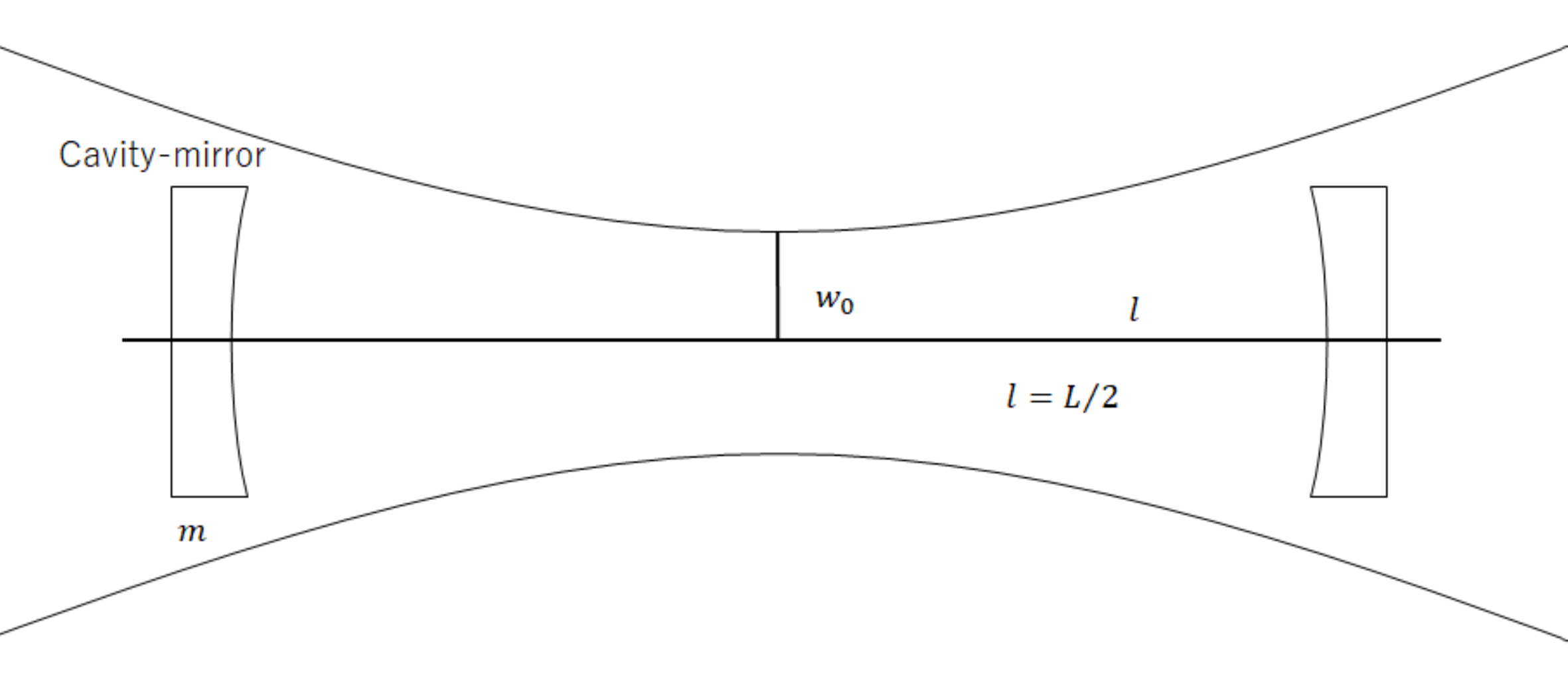}
\caption{\label{cavity}Configuration of a DECIGO's FP cavity. The horizontal line is the line connecting the centers of the mirrors. It coincides with the center of the beam axis when the cavity mirrors are aligned. The curves at the top and bottom of the figure represent the width of the laser beam.}

\end{figure}  

Figure \ref{cavity} shows the configuration of the DECIGO's FP cavity. 
The beam waist is located in the middle of each mirror. 
The beam diameter is greater as the distance from the beam waist increases. 
At each mirror, a part of the laser light passes outside the mirror. Thus, the power of reflected light is reduced. 
This is the effect of diffraction loss. 
We take this effect into consideration using the effective mirror reflectivity $r_{\mathrm{eff}}=rD^2$, where $D$ is the effect of diffraction \cite{Iwaguchi}. 
In the formula of $r_{\mathrm{eff}}$, the square of $D$ is used because we consider two effects: the leakage loss and the higher-order modes loss.
We lose the Gaussian beam of the higher-order mode because the cavity is set to resonate only with the fundamental mode. 
In this paper, we use the maximized $D$ and it is expressed as follows \cite{Ishikawa, Iwaguchi}: 

\begin{equation}
D^2_{\mathrm{max}}=1-\mathrm{exp}\left[ -\frac{2\pi}{L\lambda}R^2\right]. 
\end{equation} 

In order to maximize $D$, we set the Rayleigh length $Z_{\mathrm{R}}=l=L/2$. Thus, the laser beam radius at the cavity mirrors $r_{0}=\sqrt{2}{w_{0}}$. We calculate DECIGO’s SNR using this $D_{\mathrm{max}}$ as a function of $L$ and $R$.

\section{Derivation of Factor \textit{C\small{FTM}}}
\label{appendixB}
The purpose of this appendix is to derive $C^2_{\mathrm{\mathrm{FTM}}}$ of Section \ref{int noise}. 
$C^2_{\mathrm{\mathrm{FTM}}}$ is the ratio of the finite-test-mass power spectral density to that for the infinite-test-mass.

In order to calculate the dissipation of test mass, we use the Bessel function of order zero and order one, $J_0$ and $J_1(x)$.
In the following equations, $\xi_{\mathrm{m}}$ is the m'th zero of $J_1(x)$.
$k_{\mathrm{m}}$ is related to $\xi_{\mathrm{m}}$ by $k_{\mathrm{m}} = \xi_{\mathrm{m}} /a$, and $p_{\mathrm{m}}$ is given,

\begin{equation}
p_{\mathrm{m}}=\frac{2}{a^2 J_0^2 (\xi_{\mathrm{m}})} \int_0^a \frac{e^{-r^2/r_0^2}}{\pi r_0^2} J_0 (k_{\mathrm{m}} r)r dr,
\end{equation}
where $r$ is the distance of a point on the mirror surface from the beam spot center. $a$ is the test mass radius.

The dissipation of the finite size test mass is obtained 
\begin{equation}
\label{diss_FTM}
W_{\mathrm{diss}}=2\pi f \phi (U_0 + \Delta U) F_0^2.
\end{equation}
Here, $U_0$ is
\begin{equation}
\label{U_0}
U_0=\frac{(1-\sigma^2)\pi a^3}{E_0} \sum_{m=1}^{\infty} U_{\mathrm{m}} \frac{p^2_{\mathrm{m}} J^2_0(\xi_{\mathrm{m}})}{\xi_{\mathrm{m}}},
\end{equation}
with
\begin{equation}
U_{\mathrm{m}}=\frac{1-Q_{\mathrm{m}}^2+4k_{\mathrm{m}} h Q_{\mathrm{m}}}{(1-Q_{\mathrm{m}})^2-4k_{\mathrm{m}}^2 h^2 Q_{\mathrm{m}}},
\end{equation}
and
\begin{equation}
Q_{\mathrm{m}}=\mathrm{exp}(-2k_{\mathrm{m}} h).
\end{equation}
$\Delta U$ is expressed as
\begin{equation}
\label{deltaU}
 \Delta U=\frac{a^2}{6\pi h^3 E_0}\left[\pi^2 h^4 p_0^2 + 12\pi H^2 \sigma p_0 s + 72(1-\sigma)s^2 \right],
\end{equation}
with
\begin{equation}
p_0 = \frac{1}{\pi a^2},
\end{equation}
and
\begin{equation}
s=\pi a^2 \sum_{m=1}^{\infty} \frac{p_{\mathrm{m}} J_0 (\xi_{\mathrm{m}})}{\xi_{\mathrm{m}}^2}.
\end{equation}

Inserting Equations (\ref{diss_FTM}), (\ref{U_0}) and (\ref{deltaU}) into Equation (\ref{Levineq}), we obtain the noise power spectrum of the finite size test mass,
\begin{align}
\label{S_x^FTM}
S_{x_{\mathrm{int}}}(f)& = \frac{8k_B T}{2\pi f}\phi (U_0 + \Delta U)\\
& =C^2_{\mathrm{FTM}} \times S_{x_{\mathrm{inf}}}(f).
\end{align}

\begin{figure}[H]
\centering
\includegraphics[width=10cm]{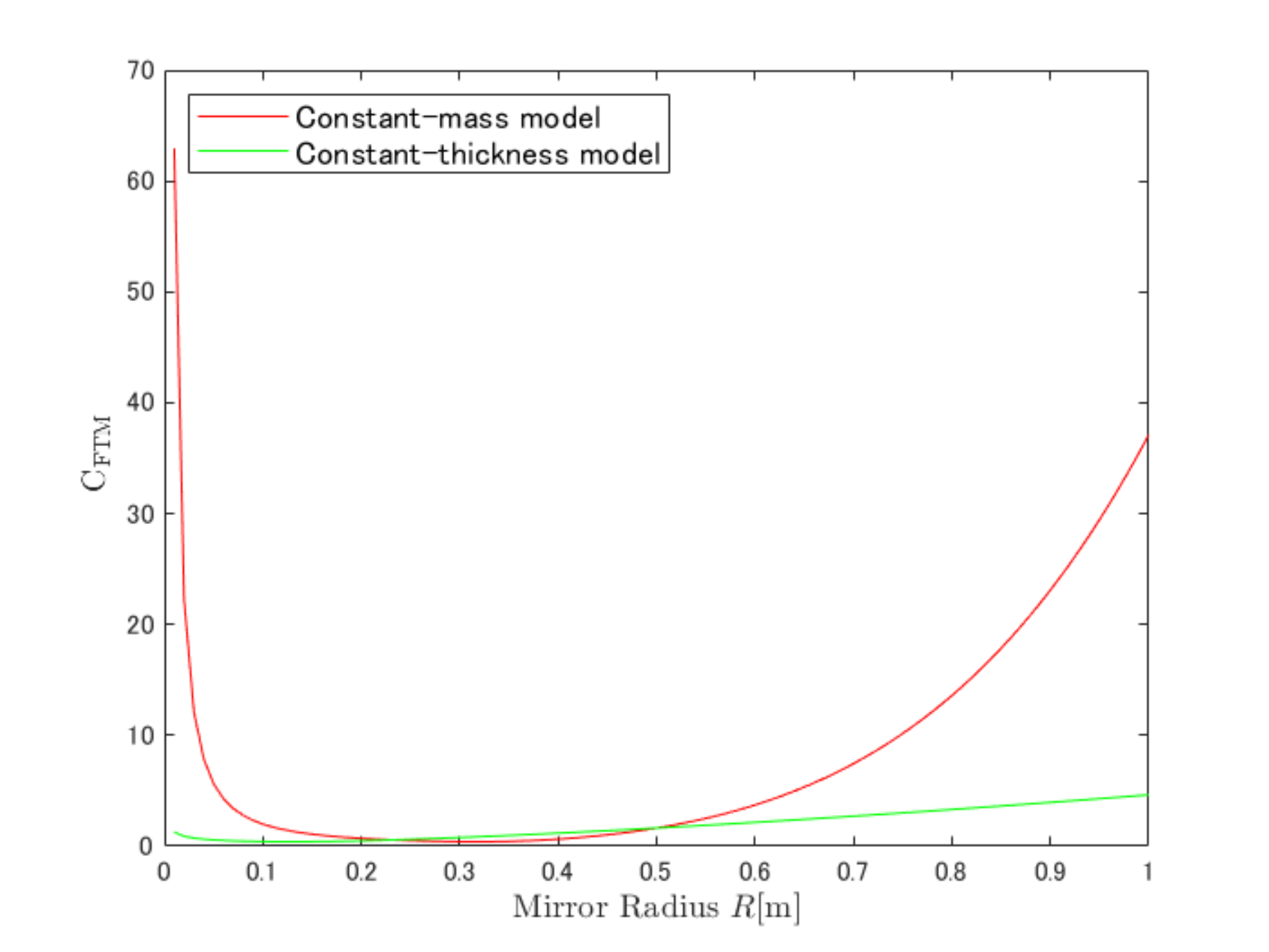}
\caption{\label{CFTM} $C_{\mathrm{FTM}}$ of constant-mirror-mass model (red line), and constant-mirror-thickness model (green line).}
\end{figure}

In our calculation, we consider up to $m=9$ for simplicity because the contribution of large $m$ is small.
Figure \ref{CFTM} shows the value of $C_{\mathrm{FTM}}$ 

{When $h/R$ is not small, the assumption in the calculation of $C_{\mathrm{FTM}}$ is incorrect. However, since the effect on the result is small, the correction when R is small (that is, $h/R$ is not small) is not considered.}

\vspace{-9pt}

\end{appendices}





\end{document}